\documentclass[prd,aps,10pt,twocolumn,nofootinbib,showpacs]{revtex4-2}

\usepackage{amssymb,amsmath}
\usepackage{xcolor}
\usepackage[utf8]{inputenc}

\usepackage{graphicx}
\usepackage{framed}
\usepackage[makeroom]{cancel}

\usepackage[makeroom]{cancel}
\usepackage[caption=false]{subfig}
\usepackage[colorlinks=true,
            citecolor=green,
            linkcolor=red,
            filecolor=cyan,
            urlcolor=magenta,
            backref=false]{hyperref}

\newcommand{\dd}{\mbox{d}}

\newcommand{\ts}[1]{{\boldsymbol{#1}}}

\begin{document}

\title{`Translation invariant' black hole: autoparallels and complete integrability}
\author{Jens Boos}
\email{jens.boos@kit.edu}
\affiliation{Institute for Theoretical Physics, Karlsruhe Institute of Technology, D-76128 Karlsruhe, Germany}

\date{June 3, 2025}

\begin{abstract}

We consider the autoparallel motion of test bodies in static, spherically symmetric spacetimes with torsion. We prove complete integrability of such motion for a wide range of off-shell geometries via four commuting autoparallel Killing vectors. Since these vectors reduce to translation generators in a certain limit, we refer to these geometries as `translation invariant.' Invoking the field equations of quadratic Poincar\'e gauge gravity we re-derive an exact Schwarzschild black hole solution endowed with a non-trivial torsion field scaling as $GM/r^2$, where $M$ denotes the ADM mass of the black hole. Studying the qualitative orbital dynamics via effective potentials we find notable discrepancies between autoparallels (straightest possible paths) and geodesics (shortest possible paths).
\end{abstract}

\maketitle

\section{Introduction}

Orbital motion has played a pivotal role in the experimental test of general relativity in the solar system. Not only does the formula for light deflection follow from such laws, they also correctly predict the perihelion shift of Mercury. More recently, the laws of relativistic orbital motion have been employed to determine the motion of a small, stellar black holes around supermassive black holes \cite{Ghez:2008ms,Gillessen:2008qv}. In such endeavors, the analytical study of orbital dynamics in black hole spacetimes is crucial. A breakthrough was reached in general relativity, when particle motion (and, later, wave equations) were shown to be separable and in some cases completely integrable due to the existence of a fundamental structure: a non-degenerate closed conformal Killing--Yano 2-form. For historical details and many key results as applied to four- and higher-dimensional black holes, we refer to \cite{Kubiznak:2008qp}; see also references therein.

Viewed from a fundamental perspective, the perturbative non-renormalizability of general relativity hints towards a fundamental incompleteness of the theory, whence modifications of general relativity are actively investigated. In the present work we would like to focus on avenues that lie beyond the Riemannian geometrical arena of general relativity and focus on theories involving the post-Riemannian geometrical feature of torsion \cite{Hehl:1976kj};\footnote{The notion of torsion arises naturally in supergravity from gauged supersymmetry \cite{Freedman:1976xh,VanNieuwenhuizen:1981ae}; in the present context, however, we will take a more minimal approach and focus on the purely gravitational sector wherein it instead arises from gauging the Poincar\'e isometry group of flat, four-dimensional spacetime, resulting in what has since been called ``Poincar\'e gauge gravity.''} for historical details and references to seminal papers we refer to the review \cite{Hehl:1994ue}; an up-to-date introduction can be found in \cite{Obukhov:2022khx}. Such gravitational theories with torsion are accompanied by a more complicated particle spectrum, including stable and ghost-free sectors \cite{Sezgin:1979zf,Sezgin:1981xs,Kuhfuss:1986rb,Karananas:2014pxa,Boos:2016cey,Blagojevic:2018dpz,Percacci:2020ddy,Barker:2024ydb}.

We would like to motivate our interest in such theories in connection to compact astrophysical objects by recalling that spacetime curvature $\mathcal{R}$ has units of inverse area, and hence we find at the surface of the Earth and at the surface of a stellar black hole, respectively,
\begin{align}
\mathcal{R} &\sim \frac{GM}{c^2 r^3} \sim 1.7 \times 10^{-23} \, \text{m}^{-2} \left(\frac{M}{M_\oplus}\right) \left( \frac{R_\oplus}{R} \right)^3 \, , \\
\mathcal{R} &\sim \frac{c^4}{G^2 M^2} \sim 4.6 \times 10^{-7} \, \text{m}^{-2} \left(\frac{M_\odot}{M}\right)^2 \, ,
\end{align}
where $\oplus$ denotes the Earth and $\odot$ denotes the sun. Conversely, spacetime torsion $\mathcal{T}$ has units of inverse length, and behaves as
\begin{align}
\mathcal{T} &\sim \frac{GM}{c^2 r^2} \sim 1.1 \times 10^{-16} \, \text{m}^{-1} \left(\frac{M}{M_\oplus}\right) \left( \frac{R_\oplus}{R} \right)^2 \, , \\
\mathcal{T} &\sim \frac{c^2}{GM} \sim 6.8 \times 10^{-4} \, \text{m}^{-1} \left(\frac{M_\odot}{M}\right) \, .
\end{align}
It is hence clear that torsion---if it exists---grows much more rapidly in the vicinity of compact astrophysical objects. This not only justifies our interest in such geometries, it necessitates the study of such phenomena in the full non-linear, strong gravity regime.

State-of-the-art weak-field experimental constraints stem from two sources. First, the so-called axial torsion piece $\mathcal{T}_\text{axial} \propto T{}_{[\mu\nu\rho]}$ couples directly to spinors \cite{Hehl:1971qi,Shapiro:2001rz}. In the weak-field regime it can hence be directly constrained via Highes--Drever experiments \cite{Lammerzahl:1997wk} to
\begin{align}
\left| \mathcal{T}_\text{axial} \right| \lesssim 10^{-15} \, \text{m}^{-1} \, .
\end{align}
This constraint fits eerily well with the na\"ive dimensional analysis for the magnitude of torsion at the surface of the Earth as presented above.

A second constraint comes from the study of the spin precession of Gravity Probe B in a weak gravitational field that is parametrized to possess torsion terms \cite{Mao:2006bb,March:2011ry,March:2011sa}, but those studies have been criticized to conflate instrinsic angular momentum (to which torsion couples) with orbital angular momentum (to which its coupling has been debated \cite{Hehl:2013qga}). For extended bodies moving in the gravitational field the Mathisson--Papapetrou--Dixon equations \cite{Mathisson:1937zz,Papapetrou:1951pa,Dixon:1964cjb} have been generalized to the presence of torsion \cite{Iosifidis:2023eom}, and the study of the motion of extended bodies reveals that they can be susceptible to the presence of spacetime torsion even in the absence of intrinsic spin-angular momentum \cite{Puetzfeld:2013rfa,Puetzfeld:2013sca,Puetzfeld:2014sja,Obukhov:2015eqa,Obukhov:2021uor}.

Previous work on integrability properties in the presence of torsion in black hole spacetimes has focused on purely axial (or ``skew'') torsion \cite{Houri:2010qc,Houri:2012eq,Agricola:2013ffa} or has been more formal in nature \cite{Batista:2015vxa,DAscanio:2019tpq,Cayuso:2019vyh}. In the present work, we want to close this gap by discussing---to the best of our knowledge---for the first time the complete integrability of the autoparallel equation of motion in a wide range of off-shell geometries with non-vanishing torsion, before specializing to the case of an exact Schwarzschild black hole solution of quadratic Poincar\'e gravity endowed with an $GM/r^2$ torsion profile. This torsion configuration, as we will show, has vanishing axial torsion, and is hence experimentally unconstrained. This paper provides the proof of integrability properties of autoparallels in such settings, and may serve as a stepping stone for deeper investigations in the future.

The remainder of this paper is organized as follows: In Sec.~\ref{sec:pgt} we will delineate the structure of Poincar\'e gauge gravity, before discussing, in detail, exact black hole solutions in that setting in Sec.~\ref{sec:bh}. Then, in Sec.~\ref{sec:auto}, we discuss the structure of the autoparallel equation for both massive and null particles, their conserved quantities, and prove their complete integrability explicitly. We summarize in Sec.~\ref{sec:conclusion}.

\section{Poincar\'e gauge gravity}
\label{sec:pgt}

In Poincar\'e gauge gravity, the fundamental commutator of covariant derivatives on a vector field is given by
\begin{align}
\left[ \nabla{}_\mu, \nabla{}_\nu \right] V{}^\rho = R{}_{\mu\nu}{}^\rho{}_\alpha V{}^\alpha - T{}_{\mu\nu}{}^\alpha\nabla{}_\alpha V{}^\rho \, ,
\end{align}
where the covariant derivative in turn is defined as
\begin{align}
\label{eq:cov-derivative}
\nabla{}_\mu X{}^\nu{}_\rho &\equiv \partial{}_\mu X{}^\nu{}_\rho + \Gamma{}^\nu{}_{\mu\alpha} X{}^\alpha{}_\rho - \Gamma{}^\alpha{}_{\mu\rho} X{}^\nu{}_\alpha \, .
\end{align}
The metricity condition $\nabla{}_\rho g{}_{\mu\nu} = 0$ then implies
\begin{align}
R{}_{\mu\nu}{}^\rho{}_\sigma &= \partial{}_\mu \Gamma{}^\rho{}_{\nu\sigma} + \Gamma{}^\rho{}_{\mu\alpha} \Gamma{}^\alpha{}_{\nu\sigma} - (\mu \leftrightarrow \nu) \, , \\
T{}_{\mu\nu}{}^\lambda &= \Gamma{}^\lambda{}_{\mu\nu} - \Gamma{}^\lambda{}_{\nu\mu} \, .
\end{align}
It is useful to split the connection into the (Riemannian) Levi-Civita part $\widetilde{\Gamma}{}^\lambda{}_{\mu\nu}$ and the (post-Riemannian) contortion tensor $K{}^\lambda{}_{\mu\nu}$ as follows:
\begin{align}
\Gamma{}^\lambda{}_{\mu\nu} &= \widetilde{\Gamma}{}^\lambda{}_{\mu\nu} + K{}^\lambda{}_{\mu\nu} \, , \\
\widetilde{\Gamma}{}^\lambda{}_{\mu\nu} &= \frac12 g{}^{\lambda\alpha} \left( \partial{}_\mu g{}_{\alpha\nu} + \partial{}_\nu g{}_{\alpha\mu} - \partial{}_\alpha g{}_{\mu\nu} \right) \, , \\
\label{eq:contortion}
K{}^\lambda{}_{\mu\nu} &= \frac12 \left( T{}_{\mu\nu}{}^\lambda - T{}_\mu{}^\lambda{}_\nu - T{}_\nu{}^\lambda{}_\mu \right)\, .
\end{align}

\subsection{Irreducible decompositions}

It is also convenient to decompose torsion and curvature into their irreducible pieces,
\begin{align}
T{}_{\mu\nu}{}^\lambda = \sum\limits_{I=1}^3 {}^{(I)}T{}_{\mu\nu}{}^\lambda \, , \quad
R{}_{\mu\nu}{}^\rho{}_\sigma = \sum\limits_{I=1}^6 {}^{(I)}R{}_{\mu\nu}{}^\rho{}_\sigma \, .
\end{align}
The curvature pieces $I=1,4,6$ correspond to the Weyl tensor, traceless Ricci tensor, and Ricci scalar, whereas the curvature pieces $I=2,3,5$ only exist in the presence of torsion. The torsion pieces $I=1,2,3$ correspond to the tensorial torsion piece, the trace torsion, and the axial (that is, totally antisymmetric) torsion piece, respectively. Their expressions, along with more details on the notation, are given explicitly in appendix \ref{sec:app}.

Note, however, that such constraints to not place any bounds on the pieces $I=1,2$ of the torsion tensor.

\subsection{Particle motion}

Within general relativity, the geodesic equation can be formulated in two equivalent ways. Either from the requirement that geodesics parallel propagate their tangent vector $u{}^\mu$,
\begin{align}
u{}^\alpha \widetilde{\nabla}{}_\alpha u{}^\mu = 0 \quad \Rightarrow \quad \frac{\dd u{}^\mu}{\dd \tau} + \widetilde{\Gamma}{}^\mu{}_{\alpha\beta} u{}^\alpha u{}^\beta = 0 \, ,
\end{align}
or from an action principle that extremizes the proper time along a given path with fixed endpoints,
\begin{align}
\delta \int\limits_a^b \dd \tau = 0 \quad \Rightarrow \quad \frac{\dd u{}^\mu}{\dd \tau} + \widetilde{\Gamma}{}^\mu{}_{\alpha\beta} u{}^\alpha u{}^\beta = 0 \, .
\end{align}
Both prescriptions yield the same equation: the geodesic equation written in terms of the Levi-Civita connection. Beyond general relativity, however, the results differ. While extremization of path length is a purely metric concept, not surprisingly even in the presence of torsion one still arrives at the geodesic equation as written above from an action principle. However, if one instead again requires parallel propagation of the curve's tangent vector one arrives instead at the autoparallel equation,
\begin{align}
u{}^\alpha \nabla{}_\alpha u{}^\mu = 0 \quad \Rightarrow \quad \frac{\dd u{}^\mu}{\dd \tau} + \Gamma{}^\mu{}_{\alpha\beta} u{}^\alpha u{}^\beta = 0 \, .
\end{align}
Note that formally this equation arises from the geodesic equation under the substitution $\widetilde{\Gamma}{}^\lambda{}_{\mu\nu} \rightarrow \Gamma{}^\lambda{}_{\mu\nu}$, or, equivalently, $\widetilde{\nabla}_\mu \rightarrow \nabla{}_\mu$. The difference of these formulations is hence given by a non-trivial torsion contribution,
\begin{align}
& \frac{\dd u{}^\mu}{\dd \tau} + \Gamma{}^\mu{}_{\alpha\beta} u{}^\alpha u{}^\beta = \frac{\dd u{}^\mu}{\dd \tau} + \widetilde{\Gamma}{}^\mu{}_{\alpha\beta} u{}^\alpha u{}^\beta + K{}^\mu{}_{\alpha\beta} u{}^\alpha u{}^\beta \, , \nonumber \\
& K{}^\mu{}_{\alpha\beta} u{}^\alpha u{}^\beta = T{}_\alpha{}^\mu{}_\beta u{}^\alpha u{}^\beta \not= 0 \, .
\end{align}
Note that the normalization property of geodesics and autoparallels remains intact,
\begin{align}
& u{}^\alpha \nabla{}_\alpha \left( u{}^\beta u{}_\beta \right) \\
&= 2u{}^\alpha u{}^\beta (\widetilde{\nabla}{}_\alpha u{}_\beta + K{}^\gamma{}_{\alpha\beta}u{}_\gamma) \\
&= 2u{}^\beta u{}^\alpha (\widetilde{\nabla}{}_\alpha u{}_\beta) + 2 K{}_{\alpha\beta\gamma} u{}^\alpha u{}^\beta u{}^\gamma = 0 \, ,
\end{align}
where the first term vanishes by assumption, and the second term vanishes due to $K{}_{(\mu\nu\rho)} = 0$. These considerations imply that geodesics keep their place in gravitational theories beyond general relativity and remain a geometrically motivated method to derive particle motion. Hence, even if autoparallel motion differs drastically from geodesic motion (which we will see explicitly in the remainder of this paper) it again does not pose stringent constraints on spacetime torsion, since autoparallels are merely non-minimally coupled versions of geodesics.

\subsection{Lagrangian and field equations}

The Lagrangian of parity-even quadratic Poincar\'e gauge gravity can be written as \cite{Obukhov:2022khx}
\begin{align}
\begin{split}
\mathcal{L} &= \frac{1}{2\kappa} ( a_0 R - 2\Lambda) \\
&\hspace{11pt}+ \frac12 \sum\limits_{I=1}^3 a_I \left( {}^{(I)}T_{\mu\nu\rho} \right)^2 \, \\
&\hspace{11pt}+ \frac{\ell_\text{Pl}^2}{2} \sum\limits_{I=1}^6 b_I \left( {}^{(I)}R_{\mu\nu\rho\sigma} \right)^2 \, .
\end{split}
\end{align}
The first line corresponds to the Einstein--Hilbert Lagrangian lifted to post-Riemannian geometries; the second line resembles the typical gauge structure of Yang--Mills theory, quadratic in the ``translation curvature'' torsion, and the third line is comprised of strong-gravity terms involving squares of the Riemann--Cartan curvature. Note that this Lagrangian has various limit points:
\begin{itemize}
\item General relativity: $a_I = 0$ for $I=1,2,3$ and $b_J = 0$ for $J=1,\dots,6$;\\[-1.6\baselineskip]
\item Quadratic gravity: $b_I = 0$ for $I=1,2,3$;\\[-1.6\baselineskip]
\item Teleparallel gravity: $a_0 = b_I = 0$ for $I=1,\dots,6$.
\end{itemize}
The field equations are obtained by variation with respect to the vielbein $e{}_i{}^\mu$ and the spin connection $\Gamma{}_i{}^\mu{}_\nu$, after which pure coordinate expressions can be substituted. The result can be written as \cite{Obukhov:2022khx}
\begin{align}
\label{eq:pg-eom}
& a_0 \left( R{}_{\mu\nu} - \frac12 R g{}_{\mu\nu} \right) + \Lambda g{}_{\mu\nu} - {q}^\text{T}_{\mu\nu} - \ell_\text{Pl}^2 \, q^\text{R}_{\mu\nu} \nonumber \\
& - (\nabla{}_\alpha - T{}_\alpha) h{}_\nu{}^\alpha{}_\mu - \frac12 T{}_{\alpha\beta\nu} h{}^{\alpha\beta}{}_\mu = \kappa T{}_{\mu\nu} \, , \\
& a_0( T{}_{\mu\nu}{}^\lambda + T{}_\mu \delta{}_\nu^\lambda - T{}_\nu \delta{}_\mu^\lambda ) - (h{}^\lambda{}_{\mu\nu} - h{}^\lambda{}_{\nu\mu}) \nonumber \\
&-2\ell_\text{Pl}^2\left[ (\nabla_\alpha - T{}_\alpha) h{}^{\lambda\alpha}{}_{\mu\nu} + \frac12 T{}_{\alpha\beta}{}^\lambda h{}^{\alpha\beta}{}_{\mu\nu} \right] = \kappa S{}^\lambda{}_{\mu\nu} \, , \nonumber
\end{align}
where we defined
\begin{align}
q{}^\text{T}_{\mu\nu} &= T{}_{\mu\alpha\beta} h{}{}_\nu{}^{\alpha\beta} - \frac14 g{}_{\mu\nu} T{}_{\alpha\beta}{}^\gamma h{}^{\alpha\beta}{}_\gamma \, , \\
h{}^{\mu\nu}{}_\lambda &= \sum\limits_{I=1}^3 a_I {}^{(I)} T{}^{\mu\nu}{}_\lambda \, , \\
q{}^\text{R}_{\mu\nu} &= R{}_{\mu\alpha}{}^{\beta\gamma} h{}{}_\nu{}^\alpha{}_{\beta\gamma} - \frac14 g{}_{\mu\nu} R{}_{\alpha\beta}{}^{\gamma\delta} h{}^{\alpha\beta}{}_{\gamma\delta} \, , \\
h{}^{\mu\nu}{}_{\rho\sigma} &= \sum\limits_{I=1}^6 b_I {}^{(I)} R{}^{\mu\nu}{}_{\rho\sigma} \, .
\end{align}
In the above, $T{}_{\mu\nu}$ is the energy-momentum tensor, and $S{}^\lambda{}_{\mu\nu}$ is the spin-angular momentum tensor that describes the spin of matter (but not its orbital angular momentum, which is captured by the in general asymmetric $T{}_{\mu\nu}$). Moreover, $\kappa$ is the gravitational constant, $\ell_\text{Pl}$ denotes a length scale at which strong gravity effects set in, and $\Lambda$ is the cosmological constant. The ten dimensionless coupling constants $a_I$ (with $I=0,\dots,3$) and $b_I$ (with $I=1,\dots,6$) determine the dynamical properties of the theory under consideration.

\section{Exact black hole solutions}
\label{sec:bh}

In the present work we will focus on a static, spherically symmetric black hole in vacuum accompanied by a non-trivial torsion background. We emphasize that this geometry is an exact solution of the underlying field equations. While this solution has been known for quite some time, and generalizations have been found including rotation, we would like to start this section by an alternative derivation of this metric from first principles. Much of the available literature employ differential form notation, which, while certainly useful, can obfuscate the character of certain computations. For this reason we will argue entirely from the coordinate basis point of view, and we will begin from an off-shell perspective (not invoking any field equations) before invoking them in several scenarios to obtain the final solution, which matches the expressions found in the literature.

\subsection{Static and spherically symmetric ansatz}

We begin with a special case of a spherically symmetric metric and the most general SO(3)-symmetric torsion,
\begin{align}
\begin{split}
\label{eq:geometry}
\dd s^2 &= -f(r)\dd t^2 + \frac{\dd r^2}{f(r)} + r^2 \dd\theta^2 + r^2\sin^2\theta\dd\varphi^2 \, , \\
T_{tr}{}^t &= T_1(r) \, , \\
T_{tr}{}^r &= T_2(r) \, , \\
T_{t\theta}{}^\theta &= T_{t\varphi}{}^\varphi = T_3(r) \, , \\
T_{r\theta}{}^\theta &= T_{r\varphi}{}^\varphi = T_4(r) \, .
\end{split}
\end{align}
It is sometimes stated that spherically symmetric torsion is specified in terms of \emph{six} independent functions \cite{Damour:2019oru} as opposed to just four \cite{Obukhov:2020hlp}. As it turns our, six non-vanishing coefficients are consistent with O(3)-symmetric; SO(3) symmetry then sets two additional coefficients to zero. Since reflection symmetry is respected by ordinary, classical matter around a central object, we hence proceed with only \emph{four} indpendent functions for the torsion field. Note also that the form of the metric ansatz is already in Schwarzschild gauge (that is, with only one independent function $f(r)$) which turns out to be sufficient to obtain black hole solutions.

We moreover consider the following set of vectors,
\begin{align}
\begin{split}
\label{eq:vectors}
\xi &= \frac{1+f}{2f} \, \partial_t + \frac{1-f}{2} \, \partial_r \, , \\
\rho_1 &= \sin\theta\cos\varphi \, \left[ \frac{1-f}{2f}  \, \partial_t + \frac{1+f}{2} \, \partial_r \right]  \\
&\hspace{11pt}+ \frac{1}{r} \cos\theta\cos\varphi \, \partial_\theta - \frac{1}{r} \frac{\sin\phi}{\sin\theta} \, \partial_\varphi \, , \\
\rho_2 &= \sin\theta\sin\varphi \, \left[ \frac{1-f}{2f} \, \partial_t + \frac{1+f}{2} \, \partial_r \right]  \\
&\hspace{11pt}+ \frac{1}{r} \cos\theta\sin\varphi \, \partial_\theta + \frac{1}{r} \frac{\cos\phi}{\sin\theta}  \, \partial_\varphi \, , \\
\rho_3 &= \frac{1-f}{2f} \cos\theta \, \partial_t + \frac{1+f}{2} \cos\theta \, \partial_r  \\
&\hspace{11pt}- \frac{1}{r}\sin\theta\,\partial_\theta \, .
\end{split}
\end{align}
The vectors are normalized
\begin{align}
1 = - \xi \cdot \xi = \rho_1 \cdot \rho_1 = \rho_2 \cdot \rho_2 = \rho_3 \cdot \rho_3 \, ,
\end{align}
with all other dot products vanishing, and for $f=1$ they reduce to the translation generators of flat spacetime,
\begin{align}
\xi = \partial_t \, , \quad
\rho_1 = \partial_x \, , \quad
\rho_2 = \partial_y \, , \quad
\rho_3 = \partial_z \, .
\end{align}

\subsection{Off-shell considerations}
With the geometric ansatz specified, let us now proceed with further considerations. In particular, we would like to link the so far unconstrained torsion functions $T_1(r), \dots, T_4(r)$ to the metric function $f(r)$. Demanding the Riemann--Cartan curvature tensor to vanish,\footnote{We would like to emphasize, though, that vanishing Riemann--Cartan curvature does \emph{not} imply vanishing Riemannian curvature that is solely derived from the Levi-Civita part of the connection.}
\begin{align}
R{}_{\mu\nu\rho\sigma} = 0 \, ,
\end{align}
we obtain exactly such constraints:
\begin{align}
\begin{split}
\label{eq:torsion}
T_1(r) &= -\frac{f'}{2f} \, , \quad
T_2(r) = -\frac{f'}{2} \, , \\
T_3(r) &= \frac{1-f}{2r} \, , \quad
T_4(r) = \frac{f-1}{2rf} \, .
\end{split}
\end{align}
Notably, this is sufficient to guarantee
\begin{align}
\nabla{}_\nu \xi{}^\mu = \nabla{}_\nu \rho{}_I^\mu = 0 \, , \quad I=1,2,3 \, .
\end{align}
This then immediately has two important implications. First, the vectors $\ts{\xi}$ and $\ts{\xi}_I$ commute under the T-bracket,
\begin{align}
\left[ \ts{\xi}, \ts{\rho}_I \right]{}^\mu_\text{T} = \left[ \ts{\rho}{}_I, \ts{\rho}_J \right]{}^\mu_\text{T} = 0 \, , 
\end{align}
which for two vectors $X{}^\mu$ and $Y{}^\mu$ is defined as
\begin{align}
[X, Y]{}^\mu_\text{T} \equiv [X, Y]{}^\mu + T{}_{\alpha\beta}{}^\mu X{}^\alpha Y{}^\beta \, .
\end{align}
The vanishing T-bracket between these vectors is the reason we refer to such off-shell geometries as `translation-invariant.' Mathematically speaking, manifolds with vanishing Riemann--Cartan curvature are referred to as Weitzenb\"ock geometries, and they belong to the class of parallelizable manifolds.

And second, the vectors satisfy 
\begin{align}
\begin{split}
\nabla{}_\mu \xi{}_\nu + \nabla{}_\nu \xi{}_\mu & = 0 \, , \\
\nabla{}_\mu \rho{}_{1 \nu} + \nabla{}_\nu \rho{}_{1 \mu} &= 0 \, , \\
\nabla{}_\mu \rho{}_{2 \nu} + \nabla{}_\nu \rho{}_{2 \mu} &= 0 \, , \\
\nabla{}_\mu \rho{}_{3 \nu} + \nabla{}_\nu \rho{}_{3 \mu} &= 0 \, .
\end{split}
\end{align}
This equation has recently been called an ``autoparallel Killing equation'' \cite{Boos:2025sld} (see also \cite{Sharif:2009vz,Peterson:2019uzn}), since it gives rise to conserved quantities under autoparallel motion,
\begin{align}
Q = u{}^\mu K{}_\mu = \text{const.} \quad \text{if} \quad \nabla_{(\mu} K{}_{\nu)} = 0 \, .
\end{align}
Four independent autoparallel Killing vectors are sufficient to guarantee the complete integrability of autoparallel motion \emph{off-shell}, which is a convenient property since it facilitates the analysis of model-independent properties of autoparallels across a wide range of gravitational theories, subject only to the constraint $R{}_{\mu\nu\rho\sigma} = 0$.

Let us conclude this off-shell section by making some remarks on possible torsion configurations based on these considerations. First, setting torsion to zero one finds
\begin{align}
T_1 = T_2 = T_3 = T_4 = 0 \quad \Rightarrow \quad f = 1 \, .
\end{align}
This shows that in this off-shell, teleparallel framework the only torsion-free solution is flat spacetime. This also follows from considering the equalities
\begin{align}
T_1 = T_2 = T_3 = T_4 \quad \Rightarrow \quad f = 1 \, , \\
T_1 = T_2 \, , \quad T_3 = T_4 \quad \Rightarrow \quad f = 1 \, .
\end{align}
The only other non-trivial pairwise equality is given by
\begin{align}
T_1 = T_4 \, , \quad T_3 = -T_4 \quad \Rightarrow \quad f = 1 - \frac{2GM}{r} \, .
\end{align}
where $2GM$ is an integration constant. Remarkably, one recovers an off-shell Schwarzschild metric with non-trivial torsion profile given by
\begin{align}
T_1 = T_4 = -\frac{GM}{r^2} \frac{1}{f} \, , \quad T_2 = -T_3 =  -\frac{GM}{r^2} \, .
\end{align}

\subsection{Going on-shell}

Let us now invoke the vacuum field equations \eqref{eq:pg-eom}. For general properties of the spherically symmetric gravitational field with torsion we refer to 

\subsubsection{General relativity}
Setting $a_I=b_I=\Lambda=0$ and only keeping $a_0 \not=0$, the field equations for \eqref{eq:geometry} immediately imply (for any $a_0$)
\begin{align}
T_1 = T_2 = T_3 = T_4 = 0 \, , \quad f = 1 - \frac{2GM}{r} \, ,
\end{align}
which is the Schwarzschild solution of general relativity.

\subsubsection{Quadratic torsion-free gravity}
As is well known, the Schwarzschild metric is also a solution of quadratic gravity in vacuum. This remains true in this framework: setting only $\Lambda = 0$ we still find a solution provided
\begin{align}
T_1 = T_2 = T_3 = T_4 = 0 \, , \quad f = 1 - \frac{2GM}{r} \, ,
\end{align}

\subsubsection{Generic Poincar\'e gauge gravity}
Allowing now the torsion to be non-zero, let us now describe an explicit solution of the field equation first found by Baekler \cite{Baekler:1981lkh}. To understand this solution better, let us first abandon the curvature constraint $R{}_{\mu\nu\rho\sigma} = 0$. We will see in a few moments how it emerges naturally and even allows for a physical interpretation. Hence, to re-iterate, we now focus on the geometry \eqref{eq:geometry} \emph{without} invoking Eq.~\eqref{eq:torsion}, or, equivalently, $R{}_{\mu\nu\rho\sigma} = 0$. The Baekler ansatz reads
\begin{align}
\begin{split}
\label{eq:baekler}
\dd s^2 &= -f(r)\dd t^2 + \frac{\dd r^2}{f(r)} + r^2 \dd\theta^2 + r^2\sin^2\theta\dd\varphi^2 \, , \\
f(r) &= 1- \frac{2GM}{r} - \frac{\Lambda_\text{eff}}{3} r^2 \, ,  \\
T_1 &= T_4 = -\frac{GM}{r^2} \frac{1}{f(r)} \, , \quad
T_2 = -T_3 = -\frac{GM}{r^2} \, .
\end{split}
\end{align}
We first note that the torsion configuration \eqref{eq:geometry} has a vanishing axial part,
\begin{align}
{}^{(3)}T{}_{\mu\nu\rho} \propto T{}_{[\mu\nu\rho]} = 0 \, ,
\end{align}
making this geometry all the more interesting for its phenomenological consequences since it is entirely unconstrained from its coupling to Dirac fermions. The Christoffel symbols are, as usual,
\begin{align}
\tilde{\Gamma}{}^t{}_{tr} &= - \tilde{\Gamma}{}^r{}_{rr} = \frac{f'}{2f} \, , \quad
\tilde{\Gamma}{}^r{}_{tt} = \frac{f f'}{2} \, , \\
\tilde{\Gamma}{}^r{}_{\theta\theta} &= -fr \, , \quad
\tilde{\Gamma}{}^r{}_{\varphi\varphi} = -fr \sin^2\theta \, , \\
\tilde{\Gamma}{}^\theta{}_{r\theta} &= \tilde{\Gamma}{}^\varphi{}_{r\varphi} = \frac{1}{r} \, , \\
\tilde{\Gamma}{}^\theta{}_{\varphi\varphi} &= -\sin\theta\cos\theta \, , \quad \tilde{\Gamma}{}^\varphi{}_{\theta\varphi} = \frac{\cos\theta}{\sin\theta} \, .
\end{align}
As a general property we note that the curvature satisfies
\begin{align}
R{}_{\mu\nu\rho\sigma} \propto (a_0 + a_1) \propto \Lambda_\text{eff} \, , \quad R = 4\Lambda_\text{eff} \, , 
\end{align}
Computing the irreducible pieces of the curvature tensor we find that
\begin{align}
{}^{(4)}R{}_{\mu\nu\rho\sigma} \propto GM \, \Lambda_\text{eff} \, , \quad
{}^{(6)}R{}_{\mu\nu\rho\sigma} \propto \Lambda_\text{eff} \, , 
\end{align}
where $I=4$ corresponds to a contribution from the tracefree Ricci tensor, and $I=6$ stems from the Ricci scalar. Conversely, all other pieces vanish,
\begin{align}
{}^{(I)}R{}_{\mu\nu\rho\sigma} = 0 ~ \text{for} ~ I=1,2,3,5 \, .
\end{align}
The quadratic curvature and torsion invariants are
\begin{align}
(R_{\mu\nu\rho\sigma})^2 = \frac{8\Lambda_\text{eff}^2}{3} \, , \quad (T{}_{\mu\nu\rho})^2 = 0 \, .
\end{align}

Let us now impose the vacuum field equations of Poincar\'e gauge gravity \eqref{eq:pg-eom} onto the geometry \eqref{eq:baekler}. We obtain the algebraic conditions
\begin{align}
2a_1 + a_2 = 0 \, , \quad
\Lambda_\text{eff} = \frac{\Lambda}{a_0} = \frac{3(a_0+a_1)}{2(b_2+b_6)\ell_\text{Pl}^2} \, .
\end{align}
The emergence of such effective cosmological constant---initially described as a ``confinement potential'' \cite{Hehl:1978yt,Baekler:1981lkh}---is well-documented in the literature \cite{Bakler:1983bm,Lee:1983af,Cembranos:2016gdt,Cembranos:2017pcs,Obukhov:2019fti}) and has recently been related to UV-finite properties of such black hole spacetimes in the context of renormalization group improvement \cite{Boos:2023xoq}.

Since we are interested in the asymptotically flat case of an isolated black hole with torsion, we will from now on set
\begin{align}
\Lambda_\text{eff} = 0 \, .
\end{align}
On the theory side, this corresponds to the following choice of coupling constants:
\begin{align} \hspace{-8pt}
a_0 = 1 \, , \quad \label{eq:couplings}
a_1 = -1 \, , \quad
a_2 = 2 \, , \quad
a_3 = -1 \, , \quad
\Lambda = 0 \, .
\end{align}
We note in passing that this choice for $\{a_1, a_2, a_3\}$ coincides with the von der Heyde model \cite{vdH:1976}, but we allow for $a_0 = 1$ whereas in der von der Heyde model one strictly has $a_0 = 0$. Our choice \eqref{eq:couplings} results in vanishing curvature everywhere, $R{}_{\mu\nu\rho\sigma} = 0$. Note that this only affects the full Riemann--Cartan curvature derived from the full affine connection $\Gamma{}^\lambda{}_{\mu\nu} = \widetilde{\Gamma}{}^\lambda{}_{\mu\nu} + K{}^\lambda{}_{\mu\nu}$. Determining, say, the quadratic Weyl-squared invariant from the Levi-Civita connection $\widetilde{\Gamma}{}^\lambda{}_{\mu\nu}$ alone, we obtain the usual Schwarzschildian expression
\begin{align}
\widetilde{C}{}_{\mu\nu\rho\sigma} \widetilde{C}{}^{\mu\nu\rho\sigma} = \frac{48 (GM)^2}{r^6} \, ,
\end{align}
precisely as in general relativity. Geometries with vanishing Riemann--Cartan curvature but non-trivial torsion are called Weitzenb\"ock geometries, and often encountered in teleparallel gravity. We wish to emphasize, however, that we work in the framework of Poincar\'e gauge gravity which manifestly allows the presence of non-trivial Riemann--Cartan curvature.

\subsection{Geodesic Killing vectors}
Since the geometry still respects spherical symmetry we suspect that the Killing vectors of the Schwarzschild geometry in general relativity are Killing vectors of this metric. To verify this, first recall that the set of four Killing vectors is given by
\begin{align}
\widetilde{\xi} &= \partial_t \, , \\
\widetilde{\rho}_1 &= \sin\varphi \, \partial_\theta + \cot\theta\cos\varphi\, \partial_\varphi \, , \\
\widetilde{\rho}_2 &= -\cos\varphi \, \partial_\theta + \cot\theta\sin\varphi \,\partial_\varphi \, , \\
\widetilde{\rho}_3 &= \partial_\varphi \, .
\end{align}
The vector $\ts{\widetilde{\xi}}$ is timelike for $r > 2GM$ (the region with which we are concerned in this work), and the remaining three vectors $\ts{\widetilde{\rho}}_A$ vectors are spacelike but not orthogonal, spanning the $SO(3)$ Lie algebra
\begin{align}
[\ts{\widetilde{\rho}}_A, \ts{\widetilde{\rho}}_B] = -\epsilon_{ABC}\ts{\widetilde{\rho}}_C \, ,
\end{align}
where $\epsilon{}_{ABC}$ is the Levi-Civita antisymmetric object, capital Latin indices are used to refer collectively to the rotational Killing vectors, and $\epsilon{}_{123} = +1$. More generally, we consider a vector $\widetilde{K}{}^\mu$ a Killing vector if both the metric and the torsion tensor are invariant under its flow,
\begin{align}
\mathcal{L}_{\widetilde{K}} g{}_{\mu\nu} &= \widetilde{K}{}^\alpha (\partial{}_\alpha g{}_{\mu\nu}) + (\partial{}_\mu \widetilde{K}{}^\alpha) g{}_{\alpha\nu} \nonumber \\
&\hspace{11pt}+ (\partial{}_\nu \widetilde{K}{}^\alpha) g{}_{\mu\alpha} \, , \label{eq:metric-killing} \\
\mathcal{L}_{\widetilde{K}} T{}_{\mu\nu}{}^\rho &= \widetilde{K}{}^\alpha (\partial{}_\alpha T{}_{\mu\nu}{}^\rho) + (\partial{}_\mu \widetilde{K}{}^\alpha) T{}_{\alpha\nu}{}^\rho \nonumber \\
&\hspace{11pt}+ (\partial{}_\nu \widetilde{K}{}^\alpha) T{}_{\mu\alpha}{}^\rho - (\partial{}_\alpha \widetilde{K}{}^\rho ) T{}_{\mu\nu}{}^\alpha \, , \label{eq:torsion-killing}
\end{align}
where $\mathcal{L}_{\widetilde{K}}$ denotes the Lie derivative along the vector field $\ts{\widetilde{K}}$. The vectors $\ts{\widetilde{\xi}}$ and $\ts{\widetilde{\rho}}_A$ satisfy Eq.~\eqref{eq:metric-killing} by construction, and direct computation for the geometry \eqref{eq:baekler} yields that they also satisfy Eq.~\eqref{eq:torsion-killing}. In that sense, $\ts{\widetilde{\xi}}$ and $\ts{\widetilde{\rho}}_A$ are Killing vectors within the torsionful Schwarzschild geometry of Poincar\'e gauge gravity. As a direct consequence, recalling the Introduction, geodesic motion described purely by $\widetilde{\Gamma}{}^\lambda{}_{\mu\nu}$ still features the same number of conserved quantities generated by $\ts{\widetilde{\xi}}$ and $\ts{\widetilde{\rho}}_A$.

\subsection{Autoparallel Killing vectors}
The autoparallel Killing vectors are defined as solutions of \cite{Sharif:2009vz,Peterson:2019uzn,Boos:2025sld}
\begin{align}
\nabla{}_\mu K{}_\nu + \nabla{}_\nu K{}_\mu = 0 \, ,
\end{align}
and their form is as given in Eq.~\eqref{eq:vectors}---this section is drastically short, since the vectors are autoparallel Killing vectors off-shell (provided only that $R{}_{\mu\nu\rho\sigma} = 0$, which the presently discussed geometry satisfies).

\subsection{Relation between Killing vectors}
We define the following object:
\begin{align}
\tilde{\rho}_I \equiv \epsilon{}_I{}^{JK} T{}_{\alpha\beta}{}^\mu \rho{}^\alpha_J \rho{}^\beta_J
\end{align}
for $I,J=1,2,3$. This makes intuitive sense, since the spatial translations, when crossed into each other with the three-dimensional flat epsilon symbol, generate rotation vectors. Conversely, defining
\begin{align}
T{}^\mu_{0J} &= \frac12 \epsilon{}^{\alpha\beta}{}_{\gamma\delta} T{}_{\alpha\beta}{}^\mu \widetilde{\xi}{}^{\,\gamma} \widetilde{\rho}^\delta_J \, , \\
T{}^\mu_{IJ} &= \frac12 \epsilon{}^{\alpha\beta}{}_{\gamma\delta} T{}_{\alpha\beta}{}^\mu \widetilde{\rho}^\gamma_J \widetilde{\rho}^\delta_J \, ,
\end{align}
we can prove the bilinear relations
\begin{align}
\frac{GM}{r} \left( \sin\theta\cos\varphi \, \ts{\xi} + \ts{\rho}_1 \right) + T_{01} + \frac{1}{r} T_{23} = 0 \, , \\
\frac{GM}{r} \left( \sin\theta\sin\varphi \, \ts{\xi} + \ts{\rho}_2 \right) + T_{02} + \frac{1}{r} T_{31} = 0 \, , \\
\frac{GM}{r} \left( \cos\theta \, \ts{\xi} + \ts{\rho}_3 \right) - T_{03} - \frac{1}{r} T_{12} = 0 \, .
\end{align}
Therefore we can offer a somewhat preliminary interpretation of the relation between the autoparallel Killing vectors and their general relativistic counterpart: the presence of torsion allows to transmute the isometry group of general relativity to an Abelian translational group of Poincar\'e gauge gravity. This is not a general result, since it seems to require parallelizability via $R{}_{\mu\nu\rho\sigma}=0$, so we do not expect it to hold true for any black holes encountered in Poincar\'e gauge gravity. It is an interesting question whether such structures also exist in rotating black holes with torsion as described in \cite{Obukhov:2019fti}.

\section{Autoparallel dynamics and complete integrability}
\label{sec:auto}

Now that the black hole geometry has been discussed in detail, we can address the conserved autoparallel quantities and discuss the general properties of autoparallel motion in such a geometry.

\subsection{Schwarzschild geodesics in general relativity}
As a point of comparison, we will briefly discuss the main features of Schwarzschild geodesics as encountered in general relativity. The constants of motion are
\begin{align}
\begin{split}
\widetilde{E} &= -g{}_{\mu\nu}\widetilde{\xi}{}^\mu u{}^\nu = \left(1 - \frac{2GM}{r}\right)\dot{t} \, , \\
\widetilde{L}_1 &= g{}_{\mu\nu}\widetilde{\rho}^\mu_1 u{}^\nu = +r^2\sin\varphi\dot{\theta} + r^2\sin\theta\cos\theta\cos\varphi\dot{\varphi} \, , \\
\widetilde{L}_2 &= g{}_{\mu\nu}\widetilde{\rho}^\mu_2 u{}^\nu = -r^2\cos\varphi\dot{\theta} + r^2\sin\theta\cos\theta\sin\varphi\dot{\varphi} \, , \\
\widetilde{L}_3 &= g{}_{\mu\nu}\widetilde{\rho}^\mu_3 u{}^\nu = r^2\sin^2\theta \dot{\varphi} \equiv \widetilde{L} \, .
\end{split}
\end{align}
Due to spherical symmetry the motion is confined to a plane which we are free to choose to lie at $\theta=\pi/2$ without loss of generality, rendering $\widetilde{L}_1 = \widetilde{L}_2 = 0$ and leaving the angular momentum magnitude captured in $\widetilde{L}_3 = \widetilde{L}$. The resulting equations of motion can be obtained by recalling that $u{}^\mu u{}_\mu = -1$ such that
\begin{align}
\dot{t} &= \frac{r\widetilde{E}}{r-2GM} \, , \\
\dot{\varphi} &= \frac{\widetilde{L}}{r^2} \, , \\
\dot{r}^2 &= \tilde{E}^2 - \widetilde{V}_\text{eff} \, , \\
\widetilde{V}_\text{eff} &= \left(1 + \frac{\widetilde{L}^2}{r^2} \right)\left(1 - \frac{2GM}{r} \right) \, .
\end{align}

\subsection{Complete autoparallel integrability}

The autoparallel conserved quantities can now be derived as dot products between the 4-velocity and the four autoparallel Killing vectors \eqref{eq:vectors} such that
\begin{align}
E &= \dot{t} - \frac{GM}{r} \left( \dot{t} + \frac{\dot{r}}{1-\frac{2GM}{r}} \right) \, , \\
P_1 &= \left[ \frac{\dd}{\dd\tau} - \frac{GM}{r^2} \left( \dot{t} - \frac{\dot{r}}{1-\frac{2GM}{r}} \right) \right] r \sin\theta\cos\varphi \, , \\
P_2 &= \left[ \frac{\dd}{\dd\tau} - \frac{GM}{r^2} \left( \dot{t} - \frac{\dot{r}}{1-\frac{2GM}{r}} \right) \right] r \sin\theta\sin\varphi \, , \\
P_3 &= \left[ \frac{\dd}{\dd\tau} - \frac{GM}{r^2} \left( \dot{t} - \frac{\dot{r}}{1-\frac{2GM}{r}} \right) \right] r \cos\theta \, .
\end{align}
This form is useful, since one can quickly verify that they are linked to the norm of the 4-momentum via
\begin{align}
\label{eq:dispersion}
g{}_{\mu\nu} u{}^\mu u{}^\nu = -E^2 + P_1^2 + P_2^2 + P_3^2 \, .
\end{align}
In the limiting case of $GM \rightarrow 0$ we recover
\begin{align}
E = \dot{t} \, , \quad
P_1 = \dot{x} \, , \quad
P_2 = \dot{y} \, , \quad
P_3 = \dot{z} \, ,
\end{align}
but note that \eqref{eq:dispersion} holds true for $GM > 0$. However, for the purposes of studying autoparallel motion we can recast the above system as expressions for $u{}^\mu$ directly,
\begin{align}
\dot{t} &= \frac{GM}{r-2GM} \left[ \sin\theta\left( P_1\cos\varphi + P_2\sin\varphi \right) + P_3 \cos\theta  \right] \nonumber \\
&\hspace{11pt} + \frac{r-GM}{r-2GM} E \, , \\
\dot{r} &= \left( 1 - \frac{GM}{r} \right) \left[ \sin\theta\left( P_1\cos\varphi + P_2\sin\varphi \right) + P_3 \cos\theta \right] \nonumber \\
&\hspace{11pt}+ \frac{GM}{r}E \, , \\
\dot{\theta} &= \frac{1}{r}\left[ \cos\theta\left( P_1\cos\varphi + P_2\sin\varphi \right) + P_3 \sin\theta \right] \, , \\
\dot{\varphi} &= \frac{1}{r\sin\theta} \left( P_2\cos\varphi - P_1\sin\varphi \right) \, .
\end{align}
As is apparent from the above, and is expected from the spherical symmetry of the problem under consideration, it is consistent to set, without loss of generality,
\begin{align}
\dot{\theta} = 0 \, , \quad \theta = \frac{\pi}{2} \, , \quad P_3 = 0 \, .
\end{align}
Note that $P_3 = 0$ is a required condition for consistency. The motion is confined to a plane, and determined by the three constants of motion $\{E, P_1, P_2\}$:
\begin{framed}
\begin{align}
\dot{t} &= \frac{(r-GM) E + GM \left( P_1\cos\varphi + P_2\sin\varphi \right)}{r-2GM} \, , \label{eq:eom-t} \\
\dot{r} &= \left( 1 - \frac{GM}{r} \right) \left( P_1\cos\varphi + P_2\sin\varphi \right) + \frac{GM}{r}E \, , \label{eq:eom-r} \\
\dot{\varphi} &= \frac{P_2\cos\varphi - P_1\sin\varphi}{r} \, , \label{eq:eom-phi} \\
& E^2 - P_1^2 - P_2^2 = \begin{cases} -1:~ \text{massive particle} \, , \\ 0:~ \text{null particle} \, . \end{cases} \label{eq:eom-constraint}
\end{align}
\end{framed}
These equations form the basis for the rest of this paper. However, a brief discussion of the conserved quantities is in order, in particular regarding the somewhat miraculous constraint \eqref{eq:eom-constraint} that resembles a flat spacetime dispersion relation. By taking the limit $r\rightarrow\infty$ we can identify $E$ as the energy of the particle at spatial infinity; the radial and angular velocities are
\begin{align}
\left . \dot{r} \right|_{r\rightarrow\infty} &= P_1\cos\varphi + P_2\sin\varphi \, , \label{eq:rdot-infinity} \\
\left. r\dot{\varphi} \right|_{r\rightarrow\infty} &= P_2\cos\varphi - P_1\sin\varphi = \widetilde{L} \, , \label{eq:phidot-infinity}
\end{align}
where we recognize the geodesically conserved angular momentum $\widetilde{L}$, which is of course \emph{not} conserved under autoparallel motion. Before going further and discussing the full three-dimensional problem, let us first study the case of radial infall, since it allows a straightforward comparison with Newtonian dynamics.

\subsection{Radial infall}
In the case of radial infall, we set $\dot{\varphi} = 0$, upon which Eq.~\eqref{eq:eom-phi} implies
\begin{align}
0 = P_2 \cos\varphi_0 - P_1\sin\varphi_0 \, .
\end{align}
Without loss of generality we set $\varphi_0 = 0$ and $P_2 = 0$, whereupon \eqref{eq:eom-constraint} then implies
\begin{align}
P_1 = \pm \sqrt{E^2 - 1} \, .
\end{align}
We will consider the negative sign, since according to Eq.~\eqref{eq:rdot-infinity} it corresponds to radially inward motion at infinity. The resulting equations of motion then simplify,
\begin{align}
\dot{t} &= \frac{(r-GM) E - GM \sqrt{E^2-1} }{r-2GM} \, , \\
\dot{r} &= \left( \frac{GM}{r} - 1 \right) \sqrt{E^2-1} + \frac{GM}{r}E \, .
\end{align}
One may verify that $\dot{r} < 0$ for all values of $r>0$ if and only if $E \leq -1$. We emphasize that this implies that radial infall towards a central mass $M>0$ is only possible if the test particle's energy is \emph{negative}, in exact opposition of what one encounters in general relativity. Moreover, given a positive energy, the condition $\dot{r} < 0$ is only possible for a finite radial range,
\begin{align}
r > r_\star(E) \equiv \left( 1 + \sqrt{\frac{E^2}{E^2-1}}\right) GM \, .
\end{align}
For any finite energy $r_\star(E) > 2GM$, and hence any massive autoparallel object with maximal energy reaches a turning point before reaching the Schwarzschild radius.

This behavior can be tracked back to the sign of the torsion coefficients. Switching the sign $T{}_{\mu\nu}{}^\lambda \rightarrow (-1)T{}_{\mu\nu}{}^\lambda$ indeed restores the general relativistic behavior. However, this geometry would no longer be a solution of the field equations of Poincar\'e gauge gravity. In particular, the spacetime curvature would be non-vanishing, $R{}_{\mu\nu\rho\sigma} \not= 0$, in stark departure to the properties of this geometry as reported in the literature and independently verified in symbolic computation software in the present work, including the existence of conserved quantities. We will address this issue in more detail in the Conclusions.

\subsection{Equatorial motion}
While the set of equations \eqref{eq:eom-t}--\eqref{eq:eom-phi} is sufficient to fully integrate the autoparallel motion, a direct comparison to general relativity is not straightforward. To that end, we consider the equation for $\dot{r}$ and utilize the normalization condition \eqref{eq:dispersion} to find an analogous expression compared to that of general relativity in terms of a radial effective potential. We find
\begin{align}
\dot{r}^2 &= E^2 - V_\text{eff} \, , \\
V_\text{eff} &= E^2 + \left[ 1 + \left( P_2\cos\varphi - P_1\sin\varphi \right)^2 \right] \, , \\
&\hspace{11pt}-\left[ E\left(1 - \frac{GM}{r}\right) + \frac{GM}{r}(P_1\cos\varphi + P_2\sin\varphi) \right]^2 \, . \nonumber
\end{align}
Notably, this effective potential is $\varphi$-dependent, unlike in the general relativistic case, and hence it is not particularly useful to characterize the motion in the purely radial sector. However, one may still utilize the positivity requirement on $\dot{r}^2$ as a constraint on $V_\text{eff}$, given a choice of constants of motion $\{E, P_1, P_2\}$ (subject to the constraint \eqref{eq:eom-constraint}, of course). Closer inspection shows that the effective potential obeys the following symmetries:
\begin{itemize}
\item $\varphi \rightarrow \varphi + \pi/2$ maps $P_1 \rightarrow P_2$ and $P_2 \rightarrow -P_1$;\\[-1.6\baselineskip]
\item $\varphi \rightarrow \varphi + \pi$ maps $P_1 \rightarrow -P_1$ and $P_2 \rightarrow -P_2$;\\[-1.6\baselineskip]
\item $E \rightarrow -E$ maps $P_1 \rightarrow -P_1$ and $P_2 \rightarrow -P_2$.
\end{itemize}
It is possible to show that $V_\text{eff} < E^2$ for all allowed values of $\{E,P_1,P_2\}$ and $r > 2GM$. This means that $\dot{r}^2 > 0$ is satisfied identically.

\begin{figure}[!htb]
    \centering
    \includegraphics[width=0.48\textwidth]{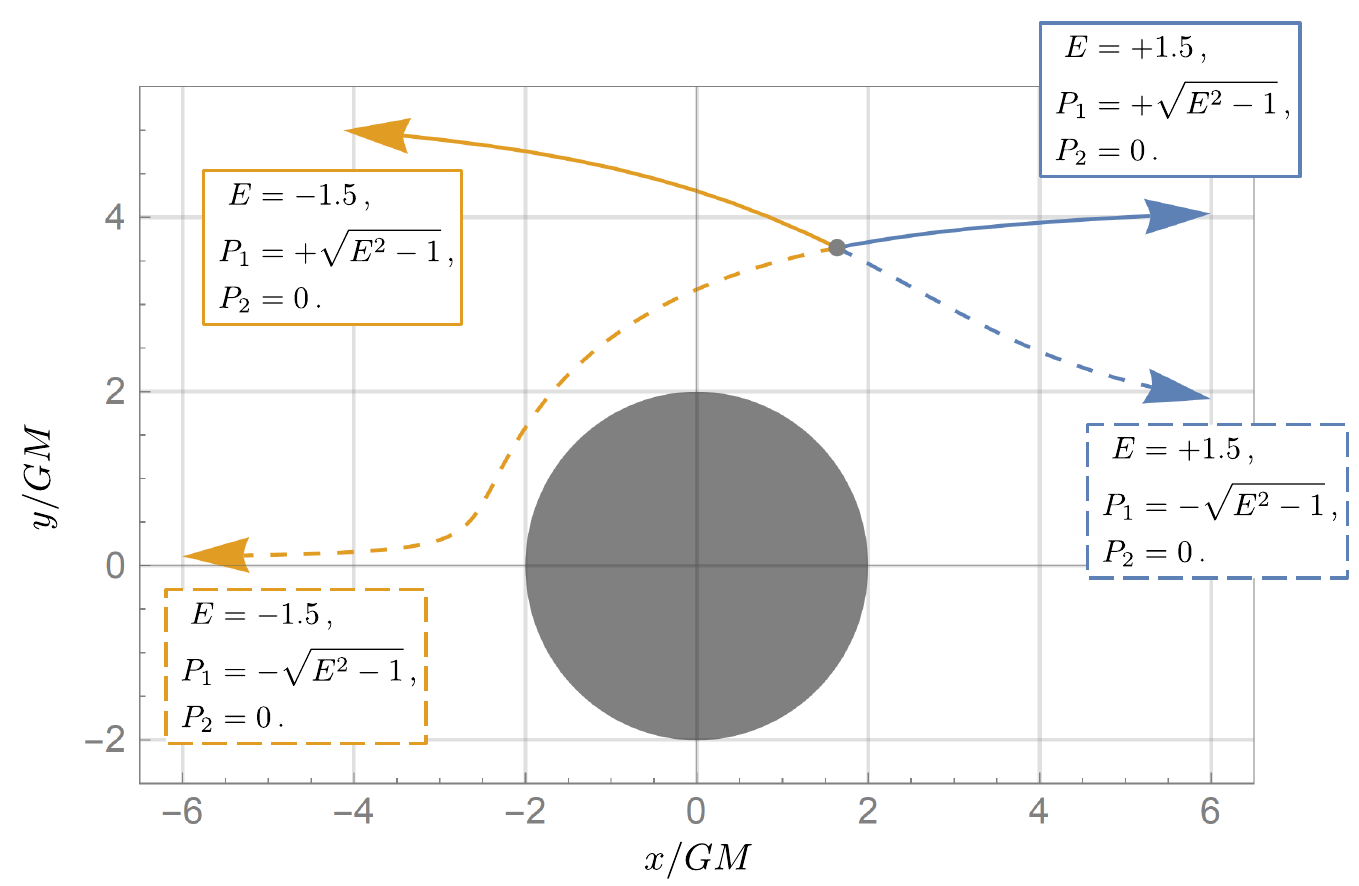} \\
    \includegraphics[width=0.48\textwidth]{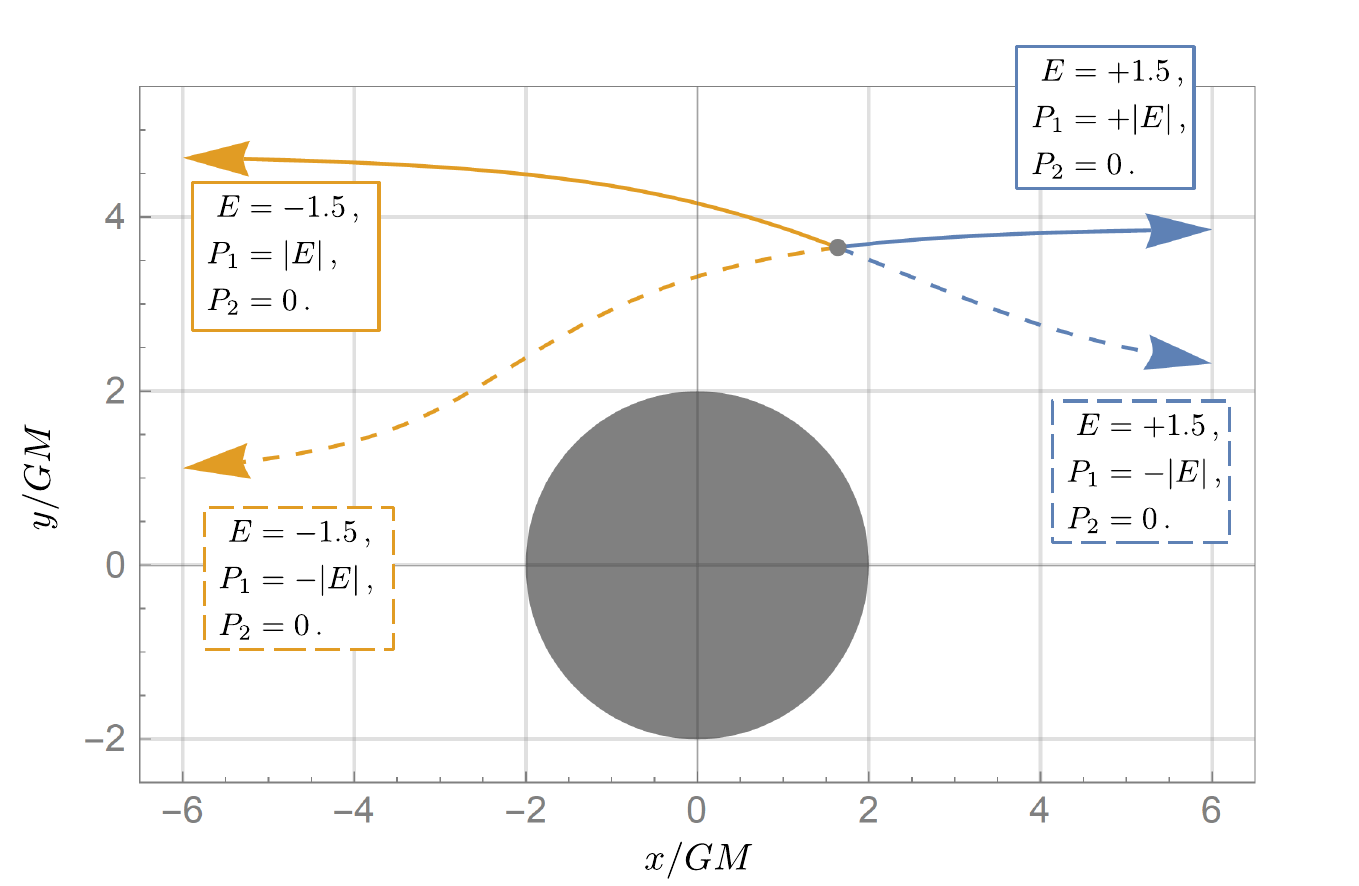} 
    \caption{Visualization of positive energy (right side) and negative energy (left side) orbits of a massive particle (top panel) and a null particle (bottom panel), in the equatorial plane. The black hole is indicated as a shaded disk. All orbits, with the exception of the negative-$E$-negative-$P_1$ ones, are repulsed by the black hole.}
    \label{fig:orbits}
\end{figure}

This seemingly anti-gravitational effect for autoparallels in this geometry can be visualized straightforwardly. Taking $P_2=0$ for simplicity, we can analyze both the timelike case and the null case,
\begin{itemize}
\item timelike: $E = \pm 1.5$ and $P_1 = \pm \sqrt{E^2-1}$;\\[-1.6\baselineskip]
\item null: $E = \pm 1.5$ and $P_1 = \pm |E|$.
\end{itemize}
Only autoparallels with sufficiently negative energy are attracted by the black hole with torsion, see Fig.~\ref{fig:orbits} for a visualization of both cases.

\subsection{Anti--Baekler geometry?}
Motivated by the previous considerations, one may instead set $M \rightarrow -M$ everywhere, including in the metric, and arrive at an autoparallel-attractive ``anti-Baekler'' geometry,
\begin{align}
\begin{split}
\label{eq:anti-baekler}
\dd s^2 &= -f(r)\dd t^2 + \frac{\dd r^2}{f(r)} + r^2 \dd\theta^2 + r^2\sin^2\theta\dd\varphi^2 \, , \\
f(r) &= 1 + \frac{2GM}{r} \, ,  \\
T_1 &= T_4 = +\frac{GM}{r^2} \frac{1}{f(r)} \, , \quad
T_2 = -T_3 = +\frac{GM}{r^2} \, .
\end{split}
\end{align}
Needless to say, the geodesics of such a geometry are repelled from it. It appears as if in this particular class of geometries, wherein the torsion is not scaled by an additional parameter, either geodesics or autoparallels must necessarily behave unphysically for positive-energy particles.

\section{Summary and conclusions}
\label{sec:conclusion}

In this work we proved for the first time the complete integrability of autoparallels for a wide class of static, spherically symmetric geometries with vanishing Riemann--Cartan curvature, $R{}_{\mu\nu\rho\sigma} = 0$. Unfortunately, the behavior of these autoparallels is rather erratic: radially infalling matter is repulsed (unless its energy is negative), and it is at present unclear if bound orbits exist (for either sign of energy)---with angular momentum not being conserved, the notion of bound orbits is seemingly different for autoparallels in the considered setting.

However, this may be due to the particularly strong torsion profile scaling as $GM/r^2$. Since there is no general Birkhoff theorem in Poincar\'e gauge gravity \cite{Obukhov:2020hlp}, it is conceivable that other spherically symmetric vacuum solutions exist, wherein the torsion profile resembles more the structure of a hair. For example, if a solution exists wherein the torsion scales less strongly as, say, $1/(GM)^2 \exp[-r/(GM)]$, the anti-gravitative effects would be much smaller. Alternatively, it is conceivable that particles move on geodesics (and not on autoparallels) which would render the conserved quantities discussed in this article useful for the study of the motion of extended bodies in such gravitational fields.

As is well known, in general relativity the geodesic equation can be derived from a point-particle limit of the field equations. At present, it is not clear whether the same is true for Poincar\'e gauge gravity, placing the autoparallel equation of motion on a less established and less geometrical footing.

It would also be insightful to study the torsionful Mathisson--Papapetrou--Dixon equation for class of geometries, perhaps even linearized in particle spin, to track the differences between orbital motion in the presence of torsion and in the absence thereof. We will leave such studies for future work.

\section{Acknowledgements}

I would like to thank Scott Hughes, his group members Lisa Drummond and Devin Becker, as well as the MIT Kavli Institute for Astrophysics for their kind hospitality during the early stages of this work, and Friedrich W. Hehl (Cologne) for many years of valuable critical correspondence on the topic of Poincar\'e gauge gravity and the physical properties of torsion. I am grateful for support as a Fellow of the Young Investigator Group Preparation Program, funded jointly via the University of Excellence strategic fund at the Karlsruhe Institute of Technology (administered by the federal government of Germany) and the Ministry of Science, Research and Arts of Baden-W\"urttemberg (Germany).

\appendix

\section{Notation}
\label{sec:app}

We begin by recalling the commutators
\begin{align}
\label{eq:fundamental-commutator-1}
[\nabla{}_\mu, \nabla{}_\nu] V{}^\rho &= R{}_{\mu\nu}{}^\rho{}_\alpha V{}^\alpha - T{}_{\mu\nu}{}^\alpha \nabla{}_\alpha V{}^\rho \, , \\
\label{eq:fundamental-commutator-2}
[\widetilde{\nabla}{}_\mu, \widetilde{\nabla}{}_\nu] V{}^\rho &= \widetilde{R}{}_{\mu\nu}{}^\rho{}_\alpha V{}^\alpha \, .
\end{align}
We define the torsionful covariant derivative $\nabla{}_\mu$ and the Levi-Civita covariant derivative $\widetilde{\nabla}_\mu$ via their action on an arbitrary (1,1)-tensor $T{}^\mu{}_\nu$ as follows:
\begin{align}
\label{eq:fundamental-cov-div-1}
\nabla{}_\mu T{}^\nu{}_\rho &\equiv \partial{}_\mu T{}{}^\nu{}_\rho + \Gamma{}^\nu{}_{\mu\alpha} T{}^\alpha{}_\rho - \Gamma{}^\alpha{}_{\mu\rho} T{}^\nu{}_\alpha \, , \\
\label{eq:fundamental-cov-div-2}
\widetilde{\nabla}{}_\mu T{}^\nu{}_\rho &\equiv \partial{}_\mu T{}{}^\nu{}_\rho + \widetilde{\Gamma}{}^\nu{}_{\mu\alpha} T{}^\alpha{}_\rho - \widetilde{\Gamma}{}^\alpha{}_{\mu\rho} T{}^\nu{}_\alpha \, .
\end{align}
By imposing the metricity condition $\nabla{}_\mu g{}_{\nu\rho} = 0$ and $\widetilde{\nabla}{}_\mu g{}_{\nu\rho} = 0$ one then readily obtains
\begin{align}
\Gamma{}^\lambda{}_{\mu\nu} &= \widetilde{\Gamma}{}^\lambda{}_{\mu\nu} + K{}^\lambda{}_{\mu\nu} \, , \\
K{}^\lambda{}_{\mu\nu} &= \frac12 \left( T{}_{\mu\nu}{}^\lambda - T{}_\mu{}^\lambda{}_\nu - T{}_\nu{}^\lambda{}_\mu \right) \, , \\
\widetilde{\Gamma}{}^\lambda{}_{\mu\nu} &= \frac12 g{}^{\lambda\alpha}\left( \partial{}_\mu g{}_{\nu\alpha} + \partial{}_\nu g{}_{\mu\alpha} - \partial{}_\alpha g{}_{\mu\nu} \right) \, .
\end{align}
In the above, $\widetilde{\Gamma}{}^\lambda{}_{\mu\nu}$ is the Levi-Civita connection, and $K{}^\lambda{}_{\mu\nu}$ is called the contortion tensor. Inserting the above relations in the commutators \eqref{eq:fundamental-commutator-1} and \eqref{eq:fundamental-commutator-2} in combination with Eqs.~\eqref{eq:fundamental-cov-div-1} and \eqref{eq:fundamental-cov-div-1} we read off the definitions of the curvature tensors $R{}_{\mu\nu}{}^\rho{}_\sigma$ and $\widetilde{R}{}_{\mu\nu}{}^\rho{}_\sigma$ as well as the torsion tensor $T{}_{\mu\nu}{}^\lambda$,
\begin{align}
R{}_{\mu\nu}{}^\rho{}_\sigma &= \partial{}_\mu \Gamma{}^\rho{}_{\sigma\nu} + \Gamma{}^\rho{}_{\alpha\mu} \Gamma{}^\alpha{}_{\sigma\nu} - \left( \mu \leftrightarrow \nu \right) \, , \\
\widetilde{R}{}_{\mu\nu}{}^\rho{}_\sigma &= \partial{}_\mu \widetilde{\Gamma}{}^\rho{}_{\sigma\nu} + \widetilde{\Gamma}{}^\rho{}_{\alpha\mu}\widetilde{\Gamma}{}^\alpha{}_{\sigma\nu} - \left( \mu \leftrightarrow \nu \right) \,  , \\
T{}_{\mu\nu}{}^\lambda &= \Gamma{}^\lambda{}_{\mu\nu} - \Gamma{}^\lambda{}_{\nu\mu} \, .
\end{align}
Torsion and curvature can be decomposed irreducibly under the Lorentz group \cite{Hehl:1994ue,Obukhov:2022khx}. For the torsion tensor there are three orthogonal pieces,
\begin{align}
T{}_{\mu\nu}{}^\lambda &= \sum\limits_{I=1}^3 {}^{(I)}T{}_{\mu\nu}{}^\lambda \, , \\
{}^{(1)} T{}_{\mu\nu}{}^\lambda &= T{}_{\mu\nu}{}^\lambda - {}^{(2)} T{}_{\mu\nu}{}^\lambda - {}^{(3)} T{}_{\mu\nu}{}^\lambda \, , \\
{}^{(2)} T{}_{\mu\nu}{}^\lambda &= \frac13 \left( T{}_\mu \delta{}^\lambda_\nu - T{}_\nu \delta{}^\lambda_\mu \right) \, , \\
{}^{(3)} T{}_{\mu\nu}{}^\lambda &= T{}_{[\mu\nu\alpha]}g{}^{\lambda\alpha} \, ,
\end{align}
where we defined the auxiliary torsion trace vector
\begin{align}
T{}_\mu = T{}_{\alpha\mu}{}^\alpha \, .
\end{align}
For the curvature tensor there are six orthogonal pieces,
\begin{align}
R{}_{\mu\nu\rho\sigma} &= \sum\limits_{I=1}^6 {}^{(I)}R{}_{\mu\nu\rho\sigma} \, , \\
{}^{(1)}R{}_{\mu\nu\rho\sigma} &= \frac12 \left( R{}_{\mu\nu\rho\sigma} + R{}_{\rho\sigma\mu\nu} \right) \\ 
&\hspace{11pt}- {}^{(4)}R{}_{\mu\nu\rho\sigma} - {}^{(6)}R{}_{\mu\nu\rho\sigma} \, , \\
{}^{(2)}R{}_{\mu\nu\rho\sigma} &= \frac12 \left( R{}_{\mu\nu\rho\sigma} - R{}_{\rho\sigma\mu\nu} \right) - {}^{(5)}R{}_{\mu\nu\rho\sigma} \, , \\
{}^{(4)} R{}_{\mu\nu\rho\sigma} &= (+1)\left( g{}_{\mu[\rho} \cancel{R}{}_{\sigma]\nu} - g{}_{\nu[\rho} \cancel{R}{}_{\sigma]\mu} \right) \, , \\
{}^{(5)} R{}_{\mu\nu\rho\sigma} &= (-1)\left( g{}_{\mu[\rho} \check{R}{}_{\sigma]\nu} - g{}_{\nu[\rho} \check{R}_{\sigma]\mu} \right) \, , \\
{}^{(6)} R{}_{\mu\nu\rho\sigma} &= \frac{1}{6} g{}_{\mu[\rho} g{}_{\sigma]\nu} R \, , \\
{}^{(3)} R{}_{\mu\nu\rho\sigma} &= -\frac{1}{4!} \epsilon^{\alpha\beta\gamma\delta}R{}_{\alpha\beta\gamma\delta} \epsilon{}_{\mu\nu\rho\sigma} \, ,
\end{align}
where we defined the traceless symmetric Ricci tensor ($\cancel{R}_{\mu\nu} = \cancel{R}_{\nu\mu}$ and $\cancel{R}_{\alpha\beta}g{}^{\alpha\beta} = 0$), the antisymmetric Ricci tensor ($\check{R}_{\mu\nu} = - \check{R}_{\nu\mu}$), as well as the Ricci scalar $R$,
\begin{align}
R{}_{\mu\nu} &= R{}_{\alpha\mu}{}^\alpha{}_\nu \, , \\
R &= R{}_{\alpha\beta} g{}^{\alpha\beta} \, , \\
\cancel{R}{}_{\mu\nu} &= R{}_{(\mu\nu)} - \frac14 R g{}_{\mu\nu} \, , \\
\check{R}_{\mu\nu} &= R{}_{[\mu\nu]} \, .
\end{align}


\begin{thebibliography}{10}

  \bibitem{Ghez:2008ms}
	A.~M.~Ghez, S.~Salim, N.~N.~Weinberg, J.~R.~Lu, T.~Do, J.~K.~Dunn, K.~Matthews, M.~Morris, S.~Yelda and E.~E.~Becklin, \textit{et al.},
	``Measuring distance and properties of the Milky Way's central supermassive black hole with stellar orbits,''
	\href{https://doi.org/10.1086/592738}{Astrophys. J.} \textbf{689}, 1044 (2008),
	\href{https://arxiv.org/abs/0808.2870}{0808.2870 [astro-ph]}.
	
  \bibitem{Gillessen:2008qv}
	S.~Gillessen, F.~Eisenhauer, S.~Trippe, T.~Alexander, R.~Genzel, F.~Martins and T.~Ott,
	``Monitoring stellar orbits around the massive black hole in the galactic center,''
	\href{https://doi.org/10.1088/0004-637X/692/2/1075}{Astrophys. J.} \textbf{692}, 1075 (2009),
	\href{https://arxiv.org/abs/0810.4674}{0810.4674 [astro-ph]}.

  \bibitem{Kubiznak:2008qp}
	D.~Kubiznak,
	``Hidden symmetries of higher-dimensional rotating black holes,''
	\href{https://arxiv.org/abs/0809.2452}{0809.2452 [gr-qc]}.

  \bibitem{Freedman:1976xh}
	D.~Z.~Freedman, P.~van Nieuwenhuizen and S.~Ferrara,
	``Progress toward a theory of supergravity,''
	\href{https://doi.org/10.1103/PhysRevD.13.3214}{Phys. Rev. D} \textbf{13}, 3214 (1976),

  \bibitem{VanNieuwenhuizen:1981ae}
	P.~Van Nieuwenhuizen,
	``Supergravity,''
	\href{https://doi.org/10.1016/0370-1573(81)90157-5}{Phys. Rept.} \textbf{68}, 189 (1981).

  \bibitem{Hehl:1976kj}
	F.~W.~Hehl, P.~Von Der Heyde, G.~D.~Kerlick and J.~M.~Nester,
	``General relativity with spin and torsion: Foundations and prospects,''
	\href{https://doi.org/10.1103/RevModPhys.48.393}{Rev. Mod. Phys.} \textbf{48}, 393 (1976).

  \bibitem{Hehl:1994ue}
	F.~W.~Hehl, J.~D.~McCrea, E.~W.~Mielke and Y.~Ne'eman,
	``Metric affine gauge theory of gravity: Field equations, Noether identities, world spinors, and breaking of dilation invariance,''
	\href{https://doi.org/10.1016/0370-1573(94)00111-F}{Phys. Rept.} \textbf{258}, 1 (1995),
	\href{https://arxiv.org/abs/gr-qc/9402012}{gr-qc/9402012}.
	
  \bibitem{Obukhov:2022khx}
	Y.~N.~Obukhov,
	``Poincar\'e gauge gravity primer,''
	\href{https://doi.org/10.1007/978-3-031-31520-6_3}{Lect. Notes Phys.} \textbf{1017}, 105 (2023),
	\href{https://arxiv.org/abs/2206.05205}{2206.05205 [gr-qc]}.

 \bibitem{Sezgin:1979zf}
	E.~Sezgin and P.~van Nieuwenhuizen,
	``New ghost-free gravity Lagrangians with propagating torsion,''
	\href{https://doi.org/10.1103/PhysRevD.21.3269}{Phys. Rev. D} \textbf{21}, 3269 (1980).	
	
  \bibitem{Sezgin:1981xs}
	E.~Sezgin,
	``Class of ghost-free gravity Lagrangians with massive or massless propagating torsion,''
	\href{https://doi.org/10.1103/PhysRevD.24.1677}{Phys. Rev. D} \textbf{24}, 1677 (1981).
	
  \bibitem{Kuhfuss:1986rb}
	R.~Kuhfuss and J.~Nitsch,
	``Propagating modes in gauge field theories of gravity,''
	\href{https://doi.org/10.1007/BF00763447}{Gen. Rel. Grav.} \textbf{18}, 1207 (1986).  

  \bibitem{Karananas:2014pxa}
	G.~K.~Karananas,
	``The particle spectrum of parity-violating Poincar\'e gravitational theory,''
	\href{https://doi.org/10.1088/0264-9381/32/5/055012}{Class. Quant. Grav.} \textbf{32}, 055012 (2015),
	\href{https://arxiv.org/abs/1411.5613}{1411.5613 [gr-qc]}.
	
  \bibitem{Boos:2016cey}
	J.~Boos and F.~W.~Hehl,
	``Gravity-induced four-fermion contact interaction implies gravitational intermediate W and Z type gauge bosons,''
	\href{https://doi.org/10.1007/s10773-016-3216-3}{Int. J. Theor. Phys.} \textbf{56}, 751 (2017),
	\href{https://arxiv.org/abs/1606.09273}{1606.09273 [gr-qc]}.
	
  \bibitem{Blagojevic:2018dpz}
	M.~Blagojevi\'c and B.~Cvetkovi\'c,
	``General Poincar\'e gauge theory: Hamiltonian structure and particle spectrum,''
	\href{https://doi.org/10.1103/PhysRevD.98.024014}{Phys. Rev. D} \textbf{98}, 024014 (2018),
	\href{https://arxiv.org/abs/1804.05556}{1804.05556 [gr-qc]}.

  \bibitem{Percacci:2020ddy}
	R.~Percacci and E.~Sezgin,
	``New class of ghost- and tachyon-free metric affine gravities,''
	\href{https://doi.org/10.1103/PhysRevD.101.084040}{Phys. Rev. D} \textbf{101}, 084040 (2020),
	\href{https://arxiv.org/abs/1912.01023}{1912.01023 [hep-th]}.

  \bibitem{Barker:2024ydb}
	W.~Barker and C.~Marzo,
	``Particle spectra of general Ricci-type Palatini or metric-affine theories,''
	\href{https://doi.org/10.1103/PhysRevD.109.104017}{Phys. Rev. D} \textbf{109}, 104017 (2024),
	\href{https://arxiv.org/abs/2402.07641}{2402.07641 [hep-th]}.

  \bibitem{Hehl:1971qi}
	F.~W.~Hehl and B.~K.~Datta,
	``Nonlinear spinor equation and asymmetric connection in general relativity,''
	\href{https://doi.org/10.1063/1.1665738}{J. Math. Phys.} \textbf{12}, 1334 (1971).

  \bibitem{Shapiro:2001rz}
	I.~L.~Shapiro,
	``Physical aspects of the space-time torsion,''
	\href{https://doi.org/10.1016/S0370-1573(01)00030-8}{Phys. Rept.} \textbf{357}, 113 (2002),
	\href{https://arxiv.org/abs/hep-th/0103093}{hep-th/0103093}.

  \bibitem{Lammerzahl:1997wk}
	C.~Lammerzahl,
	``Constraints on space-time torsion from Hughes-Drever experiments,''
	\href{https://doi.org/10.1016/S0375-9601(97)00127-8}{Phys. Lett. A} \textbf{228}, 223 (1997),
	\href{https://arxiv.org/abs/gr-qc/9704047}{gr-qc/9704047}.

  \bibitem{Mao:2006bb}
	Y.~Mao, M.~Tegmark, A.~H.~Guth and S.~Cabi,
	``Constraining torsion with Gravity Probe B,''
	\href{https://doi.org/10.1103/PhysRevD.76.104029}{Phys. Rev. D} \textbf{76}, 104029 (2007),
	\href{https://arxiv.org/abs/gr-qc/0608121}{gr-qc/0608121}.
	
  \bibitem{March:2011ry}
	R.~March, G.~Bellettini, R.~Tauraso and S.~Dell'Agnello,
	``Constraining spacetime torsion with the Moon and Mercury,''
	\href{https://doi.org/10.1103/PhysRevD.83.104008}{Phys. Rev. D} \textbf{83}, 104008 (2011),
	\href{https://arxiv.org/abs/1101.2789}{1101.2789 [gr-qc]}.

  \bibitem{March:2011sa}
	R.~March, G.~Bellettini, R.~Tauraso and S.~Dell'Agnello,
	``Constraining spacetime torsion with LAGEOS,''
	\href{https://doi.org/10.1007/s10714-011-1226-2}{Gen. Rel. Grav.} \textbf{43}, 3099 (2011),
	\href{https://arxiv.org/abs/1101.2791}{1101.2791 [gr-qc]}.
	
  \bibitem{Hehl:2013qga}
	F.~W.~Hehl, Y.~N.~Obukhov and D.~Puetzfeld,
	``On Poincar\'e gauge theory of gravity, its equations of motion, and Gravity Probe B,''
	\href{https://doi.org/10.1016/j.physleta.2013.04.055}{Phys. Lett. A} \textbf{377}, 1775 (2013),
	\href{https://arxiv.org/abs/1304.2769}{1304.2769 [gr-qc]}.

\bibitem{Mathisson:1937zz}
	M.~Mathisson,
	``Neue Mechanik materieller Systeme,''
	Acta Phys. Polon. \textbf{6}, 163-200 (1937).
	
  \bibitem{Papapetrou:1951pa}
	A.~Papapetrou,
	``Spinning test particles in general relativity. 1.,''
	\href{https://doi.org/10.1098/rspa.1951.0200}{Proc. Roy. Soc. Lond. A} \textbf{209}, 248 (1951).
  
  \bibitem{Dixon:1964cjb}
	W.~G.~Dixon,
	``A covariant multipole formalism for extended test bodies in general relativity,''
	\href{https://doi.org/10.1007/BF02734579}{Nuovo Cim.} \textbf{34}, 317 (1964).
  
  \bibitem{Iosifidis:2023eom}
	D.~Iosifidis and F.~W.~Hehl,
	``Motion of test particles in spacetimes with torsion and nonmetricity,''
	\href{https://doi.org/10.1016/j.physletb.2024.138498}{Phys. Lett. B} \textbf{850}, 138498 (2024),
	\href{https://arxiv.org/abs/2310.15595}{2310.15595 [gr-qc]}.
  
  \bibitem{Puetzfeld:2013rfa}
	D.~Puetzfeld and Y.~N.~Obukhov,
	``Unraveling gravity beyond Einstein with extended test bodies,''
	\href{https://doi.org/10.1016/j.physleta.2013.07.024}{Phys. Lett. A} \textbf{377}, 2447 (2013),
	\href{https://arxiv.org/abs/1307.3933}{1307.3933 [gr-qc]}.
  
  \bibitem{Puetzfeld:2013sca}
	D.~Puetzfeld and Y.~N.~Obukhov,
	``Equations of motion in gravity theories with non-minimal coupling: a loophole to detect torsion macroscopically?,''
	\href{https://doi.org/10.1103/PhysRevD.88.064025}{Phys. Rev. D} \textbf{88}, 064025 (2013),
	\href{https://arxiv.org/abs/1308.3369}{1308.3369 [gr-qc]}.
  
  \bibitem{Puetzfeld:2014sja}
	D.~Puetzfeld and Y.~N.~Obukhov,
	``Prospects of detecting spacetime torsion,''
	\href{https://doi.org/10.1142/S0218271814420048}{Int. J. Mod. Phys. D} \textbf{23}, 1442004 (2014).,
	\href{https://arxiv.org/abs/1405.4137}{1405.4137 [gr-qc]}.
  
  \bibitem{Obukhov:2015eqa}
	Y.~N.~Obukhov and D.~Puetzfeld,
	``Multipolar test body equations of motion in generalized gravity theories,''
	\href{https://doi.org/10.1007/978-3-319-18335-0_2}{Fund. Theor. Phys.} \textbf{179}, 67 (2015)
	\href{https://arxiv.org/abs/1505.01680}{1505.01680 [gr-qc]}.
  
  \bibitem{Obukhov:2021uor}
	Y.~N.~Obukhov and D.~Puetzfeld,
	``Demystifying autoparallels in alternative gravity,''
	\href{https://doi.org/10.1103/PhysRevD.104.044031}{Phys. Rev. D} \textbf{104}, 044031 (2021),
	\href{https://arxiv.org/abs/2105.08428}{2105.08428 [gr-qc]}.


  \bibitem{Houri:2010qc}
	T.~Houri, D.~Kubiz\v{n}\'ak, C.~Warnick and Y.~Yasui,
	``Symmetries of the Dirac operator with skew-symmetric torsion,''
	\href{https://doi.org/10.1088/0264-9381/27/18/185019}{Class. Quant. Grav.} \textbf{27}, 185019 (2010),
	\href{https://arxiv.org/abs/1002.3616}{1002.3616 [hep-th]}.  
  
	\bibitem{Houri:2012eq}
	T.~Houri, D.~Kubiz\v{n}\'ak, C.~M.~Warnick and Y.~Yasui,
	``Local metrics admitting a principal Killing--Yano tensor with torsion,''
	\href{https://doi.org/10.1088/0264-9381/29/16/165001}{Class. Quant. Grav.} \textbf{29}, 165001 (2012),
	\href{https://arxiv.org/abs/1203.0393}{1203.0393 [hep-th]}.  
  
  \bibitem{Agricola:2013ffa}
	I.~Agricola and J.~H\"oll,
	``Cones of G manifolds and Killing spinors with skew torsion,''
	\href{https://arxiv.org/abs/1303.3601}{1303.3601 [math.DG]}.
  
  \bibitem{Batista:2015vxa}
	C.~Batista,
	``Integrability conditions for Killing--Yano tensors and maximally symmetric spaces in the presence of torsion,''
	\href{https://doi.org/10.1103/PhysRevD.91.084036}{Phys. Rev. D} \textbf{91}, 084036 (2015),
	\href{https://arxiv.org/abs/1501.05029}{1501.05029 [gr-qc]}.
  
  \bibitem{DAscanio:2019tpq}
	D.~D'Ascanio, P.~B.~Gilkey and P.~Pisani,
	``Affine Killing vector fields on homogeneous surfaces with torsion,''
	\href{https://doi.org/10.1088/1361-6382/ab2774}{Class. Quant. Grav.} \textbf{36}, 145008 (2019),
	\href{https://arxiv.org/abs/1906.01694}{1906.01694 [math.DG]}.  
  
  \bibitem{Cayuso:2019vyh}
	R.~Cayuso, F.~Gray, D.~Kubiz\v{n}\'ak, A.~Margalit, R.~Gomes Souza and L.~Thiele,
	``Principal tensor strikes again: Separability of vector equations with torsion,''
	\href{https://doi.org/10.1016/j.physletb.2019.07.007}{Phys. Lett. B} \textbf{795}, 650 (2019),
	\href{https://arxiv.org/abs/1906.10072}{1906.10072 [hep-th]}.

  
  \bibitem{Damour:2019oru}
	T.~Damour and V.~Nikiforova,
	``Spherically symmetric solutions in torsion bigravity,''
	\href{https://doi.org/10.1103/PhysRevD.100.024065}{Phys. Rev. D} \textbf{100}, 024065 (2019),
	\href{https://arxiv.org/abs/1906.11859}{1906.11859 [gr-qc]}.

  \bibitem{Obukhov:2020hlp}
	Y.~N.~Obukhov,
	``Generalized Birkhoff theorem in the Poincar\'e gauge gravity theory,''
	\href{https://doi.org/10.1103/PhysRevD.102.104059}{Phys. Rev. D} \textbf{102}, 104059 (2020),
	\href{https://arxiv.org/abs/2009.00284}{2009.00284 [gr-qc]}.

  
  \bibitem{Boos:2025sld}
	J.~Boos,
	``Torsion in two dimensions: autoparallels, symmetries, and applications to black holes,''
	\href{https://arxiv.org/abs/2504.06013}{2504.06013 [gr-qc]}.  
  
  \bibitem{Sharif:2009vz}
	M.~Sharif and B.~Majeed,
	``Teleparallel Killing vectors of spherically symmetric spacetimes,''
	\href{https://doi.org/10.1088/0253-6102/52/3/11}{Commun. Theor. Phys.} \textbf{52}, 435 (2009),
	\href{https://arxiv.org/abs/0905.3212}{0905.3212 [gr-qc]}.
  
  \bibitem{Peterson:2019uzn}
	C.~Peterson and Y.~Bonder,
	``Conserved quantities in the presence of torsion: A generalization of Killing's theorem,''
	\href{https://doi.org/10.1142/S0217732320500522}{Mod. Phys. Lett. A} \textbf{35}, 2050052 (2019),
	\href{https://arxiv.org/abs/1904.12913}{1904.12913 [gr-qc]}.


  \bibitem{Baekler:1981lkh}
	P.~Baekler,
	``A spherically symmetric vacuum solution of the quadratic Poincar\'e gauge field theory of gravitation with Newtonian and confinement potentials,''
	\href{https://doi.org/10.1016/0370-2693(81)90111-8}{Phys. Lett. B} \textbf{99}, 329 (1981).

  \bibitem{Hehl:1978yt}
	F.~W.~Hehl, Y.~Ne'eman, J.~Nitsch and P.~von der Heyde,
	``Short-range confining component in a quadratic Poincar\'e gauge theory of gravitation,''
	\href{https://doi.org/10.1016/0370-2693(78)90358-1}{Phys. Lett. B} \textbf{78}, 102 (1978).  

  
  \bibitem{Bakler:1983bm}
	P.~Baekler,
	``Spherically symmetric solutions of the Poincar\'e gauge field theory,''
	\href{https://doi.org/10.1016/0375-9601(83)90179-2}{Phys. Lett. A} \textbf{96}, 279 (1983).

  \bibitem{Lee:1983af}
	C.~H.~Lee,
	``A spherically symmetric electrovacuum solution of the Poincar\'e gauge field theory of gravitation''
	\href{https://doi.org/10.1016/0370-2693(83)91137-1}{Phys. Lett. B} \textbf{130}, 257 (1983). 
  
  \bibitem{Cembranos:2016gdt}
	J.~A.~R.~Cembranos and J.~G.~Valcarcel,
	``New torsion black hole solutions in Poincar\'e gauge theory,''
	\href{https://doi.org/10.1088/1475-7516/2017/01/014}{JCAP} \textbf{01}, 014 (2017),
	\href{https://arxiv.org/abs/1608.00062}{1608.00062 [gr-qc]}.

  \bibitem{Cembranos:2017pcs}
	J.~A.~R.~Cembranos and J.~Gigante Valcarcel,
	``Extended Reissner\textendash{}Nordstr\"om solutions sourced by dynamical torsion,''
	\href{https://doi.org/10.1016/j.physletb.2018.01.081}{Phys. Lett. B} \textbf{779}, 143 (2018),
	\href{https://arxiv.org/abs/1708.00374}{1708.00374 [gr-qc]}.
  
  \bibitem{Obukhov:2019fti}
	Y.~N.~Obukhov,
	``Exact solutions in Poincar\'e gauge gravity theory,''
	\href{https://doi.org/10.3390/universe5050127}{Universe} \textbf{5}, 127 (2019),
	\href{https://arxiv.org/abs/1905.11906}{1905.11906 [gr-qc]}.  
  
  \bibitem{Boos:2023xoq}
	J.~Boos,
	``Regular black hole from a confined spin connection in Poincar\'e gauge gravity,''
	\href{https://doi.org/10.1016/j.physletb.2023.138403}{Phys. Lett. B} \textbf{848}, 138403 (2024),
	\href{https://arxiv.org/abs/2308.13017}{2308.13017 [gr-qc]}.

  \bibitem{vdH:1976}
	P.~von der Heyde,
	``Is gravitation mediated by the torsion of spacetime?,''
	\href{http://www.znaturforsch.com/aa/v31a/c31a.htm}{Z. Naturforsch.} \textbf{31a}, 1725 (1976).  
  
\end{thebibliography}
\end{document}